\documentclass[twocolumn,prl,superscriptaddress,showpacs,preprintnumbers,amsmath,amssymb]{revtex4-1}
\usepackage{graphicx}
\usepackage{color}

\newcommand{\qb}{{\bf q}}
\newcommand{\ub}{{\bf u}}
\newcommand{\vb}{{\bf v}}
\newcommand{\rb}{{\bf r}}
\renewcommand{\vec}{\bf }

\begin{document}

\title{Glass elasticity from particle trajectories}

\author{Christian L. Klix}
\affiliation{University of Konstanz, D-78457 Konstanz, Germany}
\author{Florian Ebert}
\affiliation{University of Konstanz, D-78457 Konstanz, Germany}
\author{Fabian Weysser}
\affiliation{University of Konstanz, D-78457 Konstanz, Germany}
\author{Matthias Fuchs}
\affiliation{University of Konstanz, D-78457 Konstanz, Germany}
\author{Georg Maret}
\affiliation{University of Konstanz, D-78457 Konstanz, Germany}
\author{Peter Keim}
\affiliation{University of Konstanz, D-78457 Konstanz, Germany}

\date{\today}

\begin{abstract}
Using positional data from video-microscopy of a two-dimensional colloidal system and from simulations of hard discs we determine the wave-vector-dependent elastic dispersion relations in glass. The emergence of rigidity based on the existence of a well defined displacement field in amorphous solids is demonstrated. Continuum elastic theory is recovered in the limit of long wavelengths which provides the glass elastic shear and bulk modulus as a function of temperature. The onset of a finite static shear modulus upon cooling marks the fluid/glass transition in an intuitive and unique way.
\end{abstract}


\maketitle

While fluids flow with a finite viscosity, solids respond elastically to deformation. At the glass transition, a supercooled liquid transforms into a disordered solid which possesses mechanical rigidity to shear deformations. The corresponding elastic constant is the shear modulus $\mu$ \cite{landau}. In crystalline solids, shear rigidity results from the long-ranged correlations of displacements heralding the breaking of translational invariance. The transverse acoustical displacements are the Goldstone modes, whose (squared) amplitude scales with thermal energy in the equipartition theorem.

Yet, the emergence of rigidity at the glass transition, when an amorphous solid forms, remains poorly understood. Obviously, the disorder makes it a subtle fundamental problem to apply the concepts of spontaneous breaking of translational symmetry and of Goldstone modes. Already on the macroscopic level, different predictions exist for the behaviour of the shear modulus $\mu$ when an amorphous solid forms. It has been predicted to jump discontinuously by mode coupling theory  \cite{Leutheusser1984,Goetze1985}, or to increase continuously from zero by replica theory \cite{Yoshino2010,Yoshino2012}. Such a continuous power-law increase from zero holds at the formation of a random gel, where a microscopic theory has established the link between the modulus and the Goldstone modes \cite{Zippelius2004}. Also in the theory of granular systems, critical fluctuations close to point J (which is the density where a random athermal system becomes jammed) cause a characteristic algebraic increase of the shear modulus $\mu$ from zero \cite{OHern2003}. A more recent calculation in replica theory again finds a jump in shear modulus at the glass transition and identifies the displacement field of a disordered solid, necessary to discuss elastic acoustic modes. \cite{Szamel2011}. Definitions of displacement fields in disordered solids have already been given in \cite{Goldhirsch2002}, yet the equipartition theorem could not be established in this approach.

Colloidal dispersions offer the unique possibility that the particle trajectories can be observed by video microscopy, and thus are ideally suited to study displacements microscopically. In recent work on colloidal glass \cite{Ghosh2010,Kaya2010,Chen2010}, the covariance matrix of the particle displacements was obtained and the mechanical density of states and the associated local modes were studied. But the shear modulus and the elastic dispersion relations were not obtained. Up to now only in crystalline colloidal solids, the equipartition theorem provided a route to obtain them by microscopy \cite{Gruenberg2004,Reinke2007}.\\

In this letter, we analyze the elastic properties for both an experimental 2D colloidal glass former and for simulation data of binary hard discs. We employ microscopy of displacement fluctuations, establish the equipartition theorem, and obtain the dispersion relations and elastic moduli. Our analysis is based on a quasi-equilibrium description of the non-ergodic glass state, which we derive under the assumption of a kinetic glass transition. The moduli are derived from thermally excited modes in the small wavevector limit. A sudden rise in the shear modulus marks the onset of vitrification. This proves that rigidity emerges differently in glasses than in polymeric gels and jammed granulates.\\

Here we briefly summarise our theoretical approach, which  will be given in detail elsewhere. The relation $\dot\ub_\qb(t)= \vb_\qb(t)$ is the defining relation which introduces the collective displacement field ${\bf u}_{\bf q}(t)$; $\vb_\qb(t)$ is the familiar velocity field \cite{hansen}. Fourier-modes depending on wavevector $\qb$ are considered to use translational invariance. Let time averaging be denoted by an overbar so that $\overline{r_i}$ is the average position of particle $i\in\left[1,N\right]$ during the time interval $\Delta t$, and the particle's displacement is $\ub_i(t)=\rb_i(t)- \overline{r_i}$. To order  ${\cal O}(\qb\!\cdot\! \ub_i(t)$), the field $\ub_\qb(t)$ follows by integration
\begin{equation}\label{eq0}
\ub_\qb(t) =\frac{1}{\sqrt N}\,
\sum_{i=1}^N \;
e^{ i \qb\cdot \overline{\rb_i}}\;  \ub_i(t)\; .
\end{equation}
For this result to make sense, the particle displacements $\ub_i(t)$ are required to remain bounded. For glass, this can be shown using the Zwanzig-Mori equations of motion  of the correlation functions \cite{hansen}, which are given in the supplement \cite{supplement}. They are determined by stress kernels ${\bf G}(\qb,t)$, which in the limit of  vanishing $q$ reduce to the  shear response $G(t)$  measured in macroscopic linear rheology \cite{Siebenbuerger2012}. We take  glass to be a non-ergodic state, so that the stress kernels, like density fluctuations \cite{Goetze1985,GoetzeMayr2000}, do not relax to equilibrium but take finite values at infinite time (index $\infty$),  ${\bf G}(\qb,t) \to {\bf G}_\infty(\qb)$, for $t\to\infty$. This is an idealization as it requires that the structural relaxation time $\tau_\alpha$ becomes infinite, while it (only) becomes larger than the measurement time in reality \cite{Angell1988}. Then the equations of motion lead to displacement auto-correlation functions, whose equal time value obeys  the equipartition theorem in glass \cite{supplement}
\begin{equation}\label{eq1}
\left(\langle \ub_\qb^*\;  \ub_\qb \rangle^{\rm glass}\right)^{-1} = \frac{q^2}{nk_BT} \left( {\bf G}_\infty(\qb) + \frac{\hat{\qb}\hat{\qb}}{\kappa_q^T}  \right) = \frac{1}{k_BT}\; \boldsymbol{\cal D}(\qb)\, ,
\end{equation}
where, $\kappa^T_q=S_q/(k_BTn)$ is the isothermal compressibility of the fluid with density $n$ and structure factor $S_q$.
The averaged squared displacements are given by elastic coefficients (the 'dynamical matrix' $\boldsymbol{\cal D}$), which are the frozen-in contributions in stress kernels. The superscript 'glass' indicates that averaging is done in a restricted phase space set by the glassy state. As the so obtained displacement fluctuations are ergodic, time and ensemble averages agree \cite{Kubo1957}; this was implicitly used in ansatz \eqref{eq0}. The Eigenvalues $\lambda_s(q)$ (the subscript $s$ denotes polarization) of the dynamical matrix $\boldsymbol{\cal D}$ are the 'dispersion relations' and give the elastic spectrum of the amorphous system. In high symmetry crystals in 2D as well as in amorphous solids, the elasticity tensor possesses only two independent elements. These are the Lam\'{e} coefficients $\mu$ and $\lambda$ of continuum elasticity theory.  In two dimensions, $\mu$ gives the shear and $\mu + \lambda$ the bulk modulus, extracted from the transverse and longitudinal displacements  $u_\perp(\vec q)$ and $u_\parallel(\vec q)$ in the long wavelength limit
\begin{eqnarray}
  \frac{(2\mu + \lambda)}{nk_BT}  & = & \lim_{\vec q \rightarrow 0} \left[q^2 \langle | u_\parallel(\vec q) |^2 \rangle^{\rm glass}\right]^{-1} \, ,\label{elastic_constants1} \\
  \frac{\mu}{nk_BT} & = & \lim_{\vec q \rightarrow 0} \left[ q^2 \langle | u_\perp(\vec q) |^2 \rangle^{\rm glass} \right]^{-1} \, .\label{elastic_constants2}
\end{eqnarray}

\begin{figure}[ht]
\includegraphics[width=.95\linewidth]{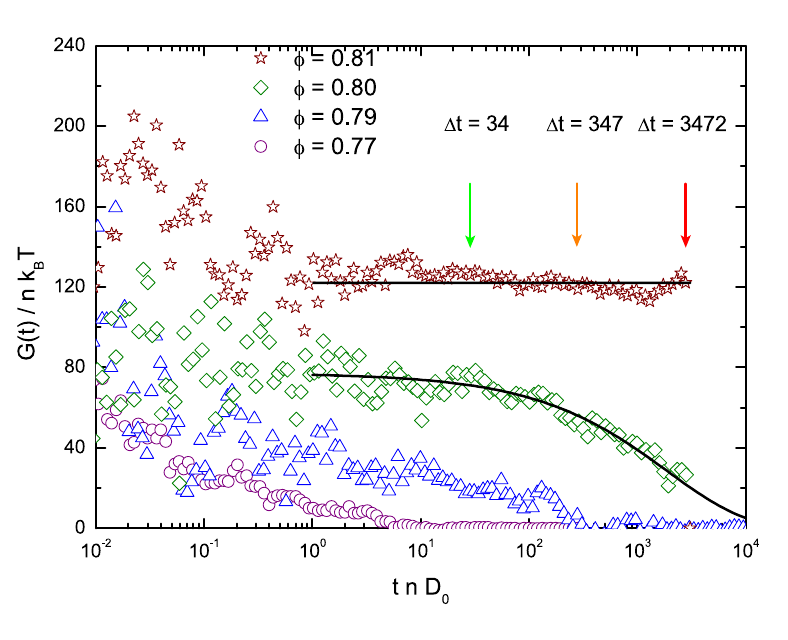}
\caption{Shear modulus $G(t)$ from simulations of a glass forming mixture of hard disks for different packing fractions; $D_0$ is the dilute diffusion constant. Arrows indicate the length $\Delta t$ of the trajectories, used to calculate the displacement fluctuations used in Fig.~\ref{fig02}. At $\phi = 0.81$, a horizontal bar marks the plateau value $G_\infty = 124\ nk_BT$. At $\phi = 0.80$, a Kohlrausch law $G_\infty e^{-(t/\tau_\alpha)^\beta}$, is fitted to the final decay; $G_\infty=77n k_BT$,  $\tau_\alpha=1900 nD_0$, and $\beta=0.60$.} \label{fig01}
\end{figure}
In order to validate our prediction in Eq.~\eqref{eq1} that the shear modulus $G_\infty=G(t\to\infty)$ derived from frozen-in stress correlations equals the one from static displacement fluctuations, $\mu=G_\infty$, we performed Brownian dynamics simulations of a binary mixture of hard discs in two dimensions \cite{supplement,Weysser2011}. Figure \ref{fig01} shows the time-dependent shear stress correlation function $G(t)$  for (area) packing fractions $\phi$ close to the glass transition. The mode coupling glass transition was determined as $\phi_c\approx 0.796$ \cite{Weysser2011}. The $G(t)$ exhibit a two-step relaxation process with a finite amplitude of the final (so called $\alpha$-) process. It only decays for times comparable to the structural relaxation time $\tau_\alpha$. Conventionally, it is obtained from fitting e.g.~a Kohlrausch law to the final decay \cite{Angell1988,kohlrausch}. When, upon raising $\phi$, the time  $\tau_\alpha$ increases beyond the observation window, a finite 'frozen-in' amplitude remains which is $G_\infty$. Figure \ref{fig01} shows that at intermediate times in the supercooled fluid, and in the glass, the shear elastic constant is finite.\\

\begin{figure}[ht]
\includegraphics[width=.95\linewidth]{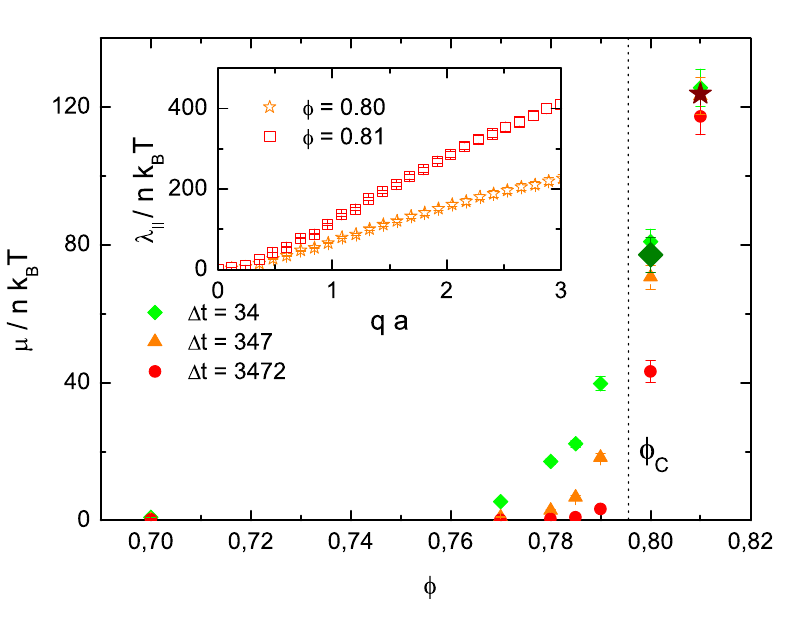}
\caption{Shear moduli $\mu$ (in $nk_BT$) from simulation as function of area fraction $\phi$. With increasing trajectory length ($\Delta t$ given in units of $nD_0$), $\mu$ converges towards zero  in the fluid. Crossing $\phi_\mathrm{c}$, $\mu$ jumps to a finite value for $\Delta t<\tau_\alpha$. $G_\infty = 124$ at $\phi = 0.81$ and  $G_\infty = 77$ at $\phi = 0.80$ (dark star and green rhombus from Fig.~\ref{fig01}) agree within  errors with $\mu$ from Eq.~(\ref{elastic_constants2}). The inset shows the  transverse glass dispersion relations.}\label{fig02}
\end{figure}

Yet the simulations in Fig.~\ref{fig01} are very demanding so that earlier ones remained inconclusive on $G_\infty$ \cite{Barrat1988}, which advocates the more efficient method to correlate the displacement fluctuations. Quite intuitively, we use the time-averaged center of the trajectory as 'equilibrium' site of a particle. Its displacement is determined relative to this position as function of time \cite{Keim2004}. This goes beyond 3D crystals, where lattice sites can be used. The latter is unfeasible in amorphous solids as well as in 2D crystals where strict long range translational order is spoiled due to Mermin-Wagner fluctuations \cite{Mermin1968}. Arrows in Fig.~\ref{fig01} indicate the lengths $\Delta t$ of different trajectories used to calculate the shear modulus $\mu$ from the displacement fluctuations according to Eq.~(\ref{elastic_constants2}). The expression for $\overline{r_i}$ holds as long as $\Delta t$ is smaller than the relaxation time $\tau_\alpha$. The results for $\mu$ are shown in Fig.~\ref{fig02} for different $\Delta t$ and $\phi$. The comparison for the  states at $\phi=0.80$, and $0.81$ shows that $\mu$ calculated from displacement fluctuations agrees  with $G_\infty$ calculated from the stress autocorrelation function. This verifies our theory. The modulus $\mu$ takes finite values as long as the structural relaxation time $\tau_\alpha$ exceeds $\Delta t$, as can be seen from Fig.~\ref{fig01}. Approaching the mode coupling glass transition at $\phi_c$ from above, $\mu$ softens but stays finite in Fig.~\ref{fig02}. Note that the time-scale in Fig.~\ref{fig01} is logarithmic: Because of the equidistant sampling used to measure $\langle|{\bf u}_\perp({\bf q})|^2\rangle$, the final part of the time signal dominates. This explains that in Fig.~\ref{fig02} for $\phi=0.80$ the measurement using $\Delta t=3472$  is below the shorter ones. The final $\alpha$-process already affects this measurement. While this decay lacks a theoretical description, for the present context it suffices that it does not affect $G_\infty$ \cite{Adichtchev2002}. On the fluid side, the modulus drops to very small values for $\Delta t$ longer than the timescale of short time correlations. Because  Eq.~(\ref{elastic_constants2}) measures a squared quantity, it can only give a positive value, which, however, converges to zero with increasing $\Delta t$ in the fluid.

After having validated our method, we apply it to the experimental 2D colloidal glass. The system consists of a binary mixture of super-paramagnetic polystyrene spheres confined to a flat water-air interface \cite{Ebert2009}. The species A (diameter $\sigma = 4.5\,\mu m$) and B ($\sigma = 2.8\,\mu m$) have a relative concentration of $\xi = N_B/(N_A+N_B)\approx 45\,\%$ where $N_A$ and $N_B$ are the number of particles of both species in the field of view. An external magnetic field $H$, perpendicular to the interface, lets us control the particle interactions \textit{in situ}. This is expressed by the dimensionless coupling parameter
\begin{equation}\label{gamma}
\Gamma = \frac{\mu_0}{4\pi} \cdot \frac{H^2 \cdot (\pi n)^{3/2}}{k_BT}(\xi \cdot \chi_B + (1-\xi)\cdot \chi_A)^2 \, ,
\end{equation}
which acts as an inverse temperature. $n$ denotes the 2D number density computed via Voronoi tessellation. $\chi_{A,B}$ is the susceptibility of species A, B. Video microscopy and digital image analysis provide the position of individual particles as function of time.\\
\begin{figure}[ht]
	\includegraphics[width=.95\linewidth]{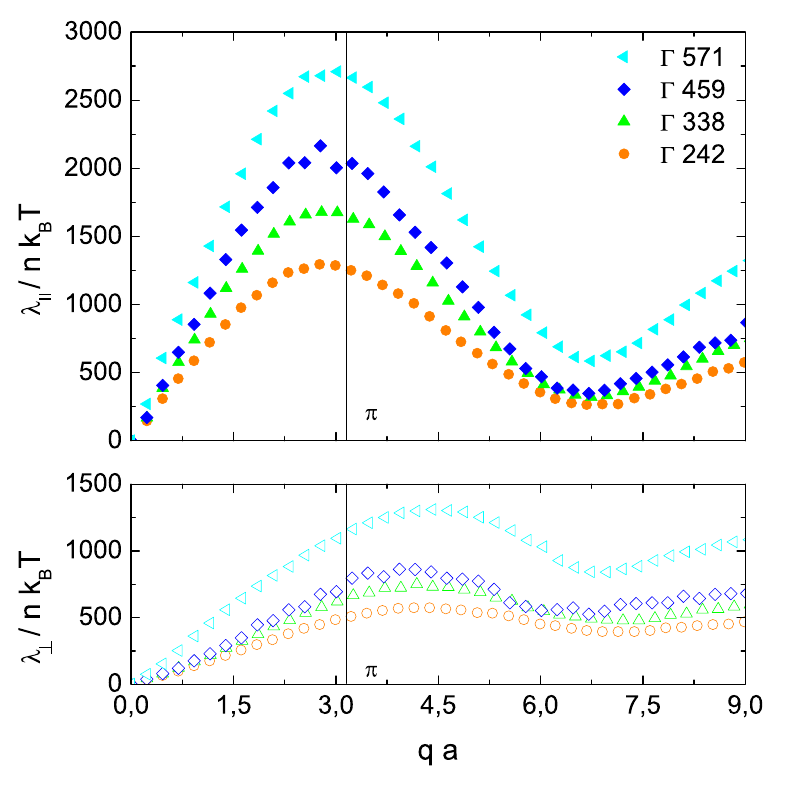}
\caption{The longitudinal (up) and transversal (bottom) dispersion relation of 2D colloidal glass for different interaction parameters  $\Gamma$ (inverse temperatures) as extracted from the displacements of particles from their time-averaged positions. Filled and empty symbols represent longitudinal and transverse polarizations, respectively.}\label{fig03}
\end{figure}

In Fig.~\ref{fig03}, the measured dispersion relations are shown for different interaction parameters $\Gamma$. Filled and empty symbols represent spring constants for longitudinal and transverse modes, respectively. The existence of transverse elastic modes signals the solid. The small noise in the measured curves is remarkable and unexpected based on the complicated eigenvalue- and eigenvector structure of the covariance-matrices obtained from the instantaneous displacements in real space \cite{Ghosh2010, Kaya2010, Chen2010}. The Fourier-transformed displacements introduced in \eqref{eq0} apparently are a good set of modes. Their dispersion correlates with the mean particle separation $a$ defined from the density $a^2 n=1$ revealing to the short-range order of the glass. The growing amplitude of the curves for increasing $\Gamma$ reflects the expected stiffening of the system upon cooling.\\
\begin{figure}[ht]
\includegraphics[width=.95\linewidth]{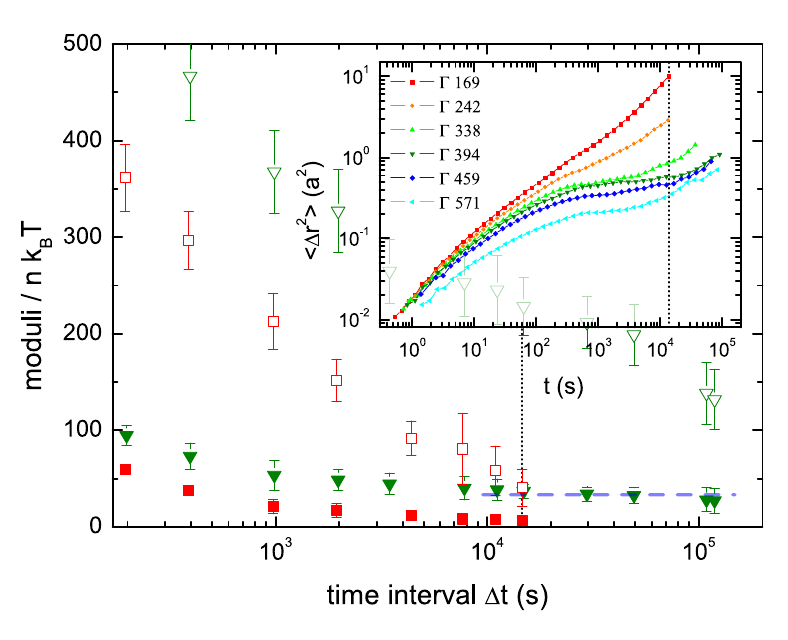}
\caption{The shear $\mu$ (filled symbols) and bulk modulus $\lambda+\mu$ (empty symbols) depending on sampling time $\Delta t$ are shown for a glassy ($\Gamma=394$, triangles) and a fluid state ($\Gamma=169$, squares). For short trajectories $\Delta t$, the moduli increase. On long timescales, $\mu$ (green filled triangles) does not change in the glass (blue dashed line gives best fit), whereas it has decayed for the fluid (red filled squares). In the glass, $\lambda+\mu$ decreases at long times when $\Delta t\approx\tau_\alpha$. The inset shows the mean squared displacement for different temperatures. Dashed vertical line indicate the sampling time $\Delta t = 14400$.
}\label{fig04}
\end{figure}

\begin{figure}[!t]
	\includegraphics[width=.95\linewidth]{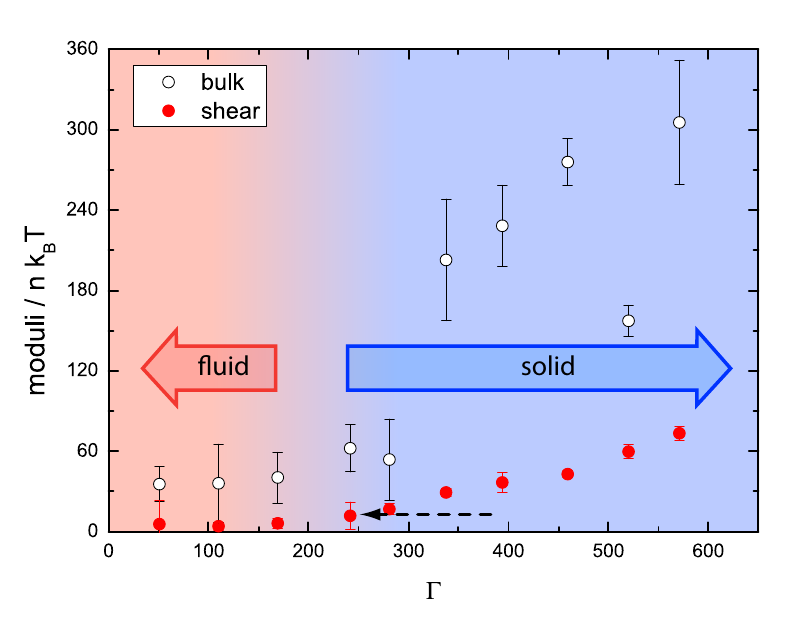}
\caption{At $\Gamma \approx 200$, we find an onset of the shear modulus (red circles) marking the temperature where the system starts to undergo vitrification. Cooling the system below $T_g$ (increasing $\Gamma$), growing spring constants lead to an increase in stiffness. The arrow indicates the value of a mode coupling calculation at its transition temperature.}\label{fig05}
\end{figure}

From the dispersion relations, the moduli can be obtained via Eqs.~(\ref{elastic_constants1},\ref{elastic_constants2}). We chose an intermediate regime ($0.8 < q a < 2.0$) for the extrapolation $q \to 0$ where the data for linear regression fits best \cite{Gruenberg2004,comment}. As the time interval  $\Delta t$ enables us to differentiate between high frequency and low frequency moduli, we discuss the dependence on $\Delta t$ first. Reducing $\Delta t$ (corresponding to high frequencies), Fig.~\ref{fig04} clearly indicates the expected growth of the shear modulus for both typical fluid ($\Gamma = 169$) and solid ($\Gamma = 394$) state. These results are consistent with \cite{Jeong1987}. At the same time, we see that for long sampling times $\Delta t$, its exact value makes no significant difference for computing the shear modulus. This is marked by a plateau (blue dashed line in Fig.~\ref{fig04}). The plateau is more stable than the one in the MSD \cite{Koenig2005,Mazoyer2009}, which is already affected by the $\alpha$-process at a correlation time $\tau_\alpha^{\rm MSD}$ of about a few $10^5~s$  (inset of Fig.~\ref{fig04}). The plateau in the shear modulus, however, ranges from $10^4\;s$ up to $10^5\;s$ (which is the longest experimental sampling time we measured). In Figure~\ref{fig04} and \ref{fig05} the bulk modulus is included. It stays finite in the fluid phase for large $\Delta t$ as expected. But since the amplitude square of the longitudinal mode is ten times smaller compared to the transverse one, it is more strongly affected by structural relaxations at $\tau_\alpha$ (e.g. aging) and/or Mermin-Wagner fluctuations \cite{Mermin1968} known from 2D crystals and also by a finite particle resolution \cite{comment} (see the error bars of Fig.~\ref{fig05}). Therefore we focus on the shear modulus in the following.

Fig.~\ref{fig05} displays the measured temperature dependence of the moduli \cite{comment_II}. As expected, the shear modulus $\mu$ is zero in the fluid phase (first three data points). As we further cool down our sample, we find an onset of $\mu$, indicating the beginning of vitrification where the system becomes rigid (indicated by the shading). The magnitude of $\mu$ compares well to a mode coupling calculation $\mu=13nk_BT$ (dashed arrow in Fig.~\ref{fig05}) \cite{Seyboldt2011}. With the given temperature resolution of the experimental data we can not decide unambiguously if the shear modulus jumps discontinuously (as we did with the simulation data). But Fig.~\ref{fig05} clearly marks the region of vitrification wich was unknown for 2D amorphous systems with long range particle interaction.\\

In conclusion, we have shown that video microscopy allows us to obtain the elastic dispersion relation of an amorphous solid directly from the spatially Fourier transformed displacement fields. The classical equipartition theorem holds and in the long wavelength limit we can extract the elasticity from the dispersion relation. A finite quasi-equilibrium shear modulus marks the onset of vitrification in the experimental 2D system. With simulation data we were able to show that this transition is discontinuous (unlike in granular systems or gels). Our work establishes microscopy of displacements as novel spectroscopic technique in supercooled liquids and glasses. This is of especial interest in soft matter systems where the classical definition using a viscosity $\eta > 10^{13}$~Poise does not hold.\\

\begin{acknowledgments}
P.K. gratefully acknowledge financial support of the Young Scholar Fund, University of Konstanz.
\end{acknowledgments}



\begin{figure*}
	\includegraphics[width=1.\textwidth]{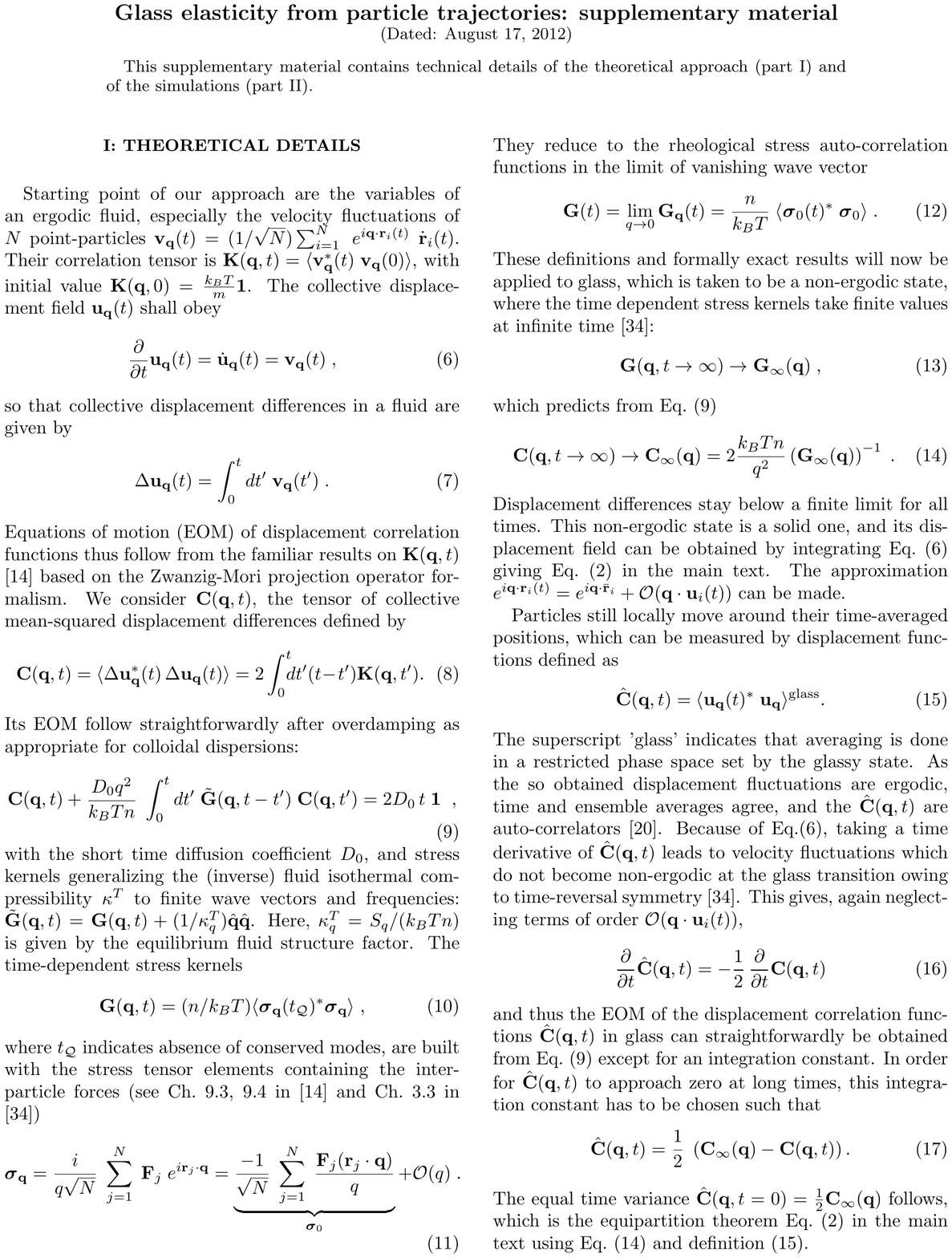}
\end{figure*}
\begin{figure*}
	\includegraphics[width=1.\textwidth]{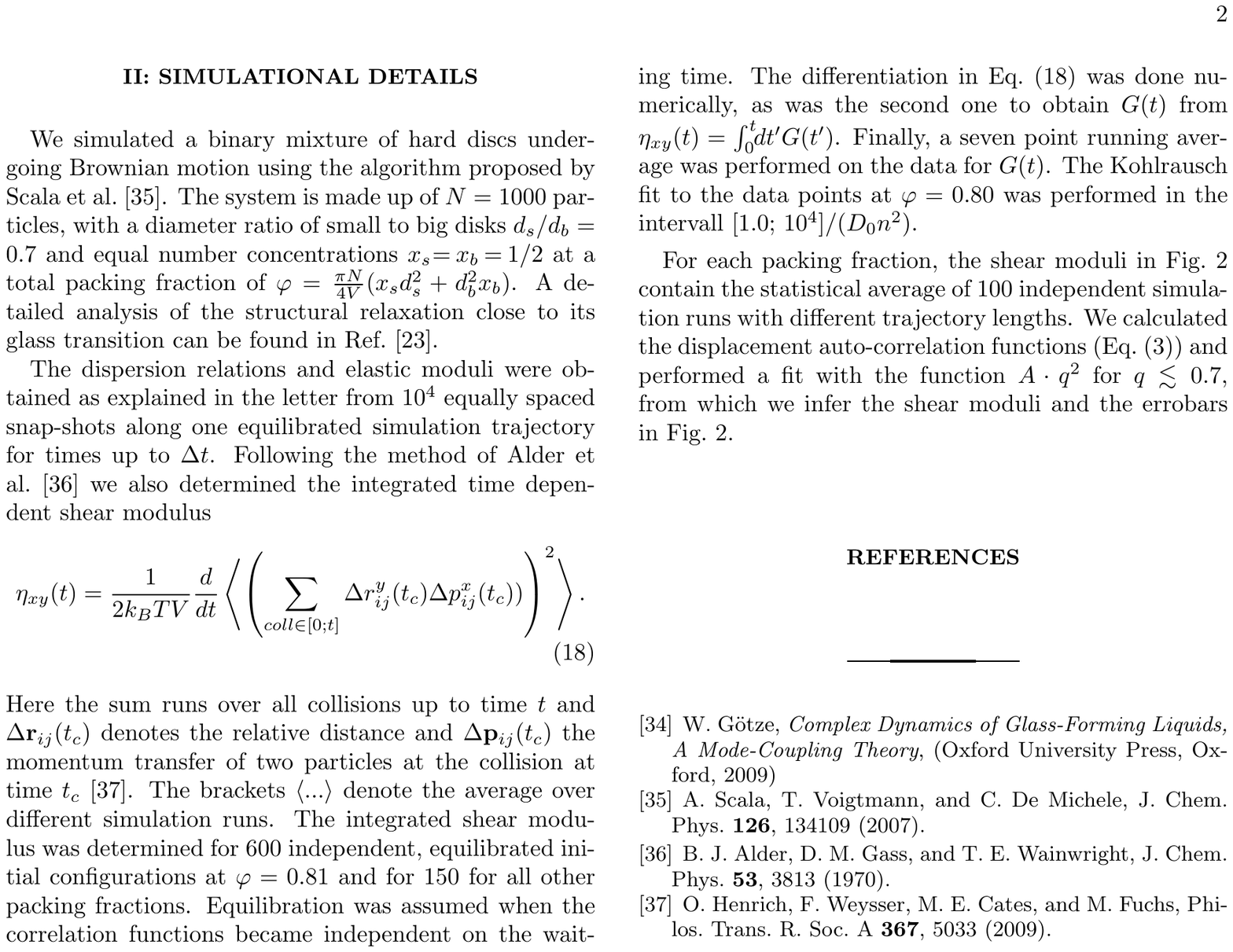}
\end{figure*}


\begin{thebibliography}{99}
\bibitem{comment} The error bars for the simulations in Fig.\;2 measure a statistical error. Since we can not prepare identical representations (with identical histories) for the experimental system the error bars of Fig.\;4, and\;5 stem from the linear regression of the extrapolation $q \to 0$ and the systematic error due to finite resolution in particle position (about $100\;nm$).
\bibitem{comment_II} $\Delta t$ was chosen to be $\Delta t\approx 14400\,s$  for the experimental data. It assures that we measure the glassy elasticity not in the high frequency region but at times shorter than $\tau_\alpha$.
\end{thebibliography}
\end{document}


\bibliographystyle{plainnat}

\title{Glass elasticity from particle trajectories: supplementary material}

\date{\today}

\begin{abstract}
This supplementary material contains technical details of the theoretical approach (part I) and of the simulations (part II).
\end{abstract}

\maketitle
\addtocounter{equation}{5}
\addtocounter{figure}{5}

\section{{\bf I:} Theoretical details}

Starting point of our approach are the variables of an ergodic fluid, especially
the velocity fluctuations  of $N$ point-particles  $\vb_\qb(t)=(1/\sqrt{N}) \sum_{i=1}^N \; e^{ i \qb\cdot \rb_i(t)}\; \dot\rb_i(t)$.
 Their correlation tensor is ${\bf K}(\qb,t) = \langle \vb_\qb^*(t)\; \vb_\qb(0) \rangle $, with initial value  ${\bf K}(\qb,0)= \frac{k_BT}{m} \boldsymbol{1}$.
The collective displacement field $\ub_\qb(t)$  shall obey
\beq{e6m}
\frac{\partial}{\partial t} \ub_\qb(t)  =  \dot \ub_\qb(t)  = \vb_\qb(t)\; ,
\eeq so that collective displacement differences in a fluid are given by
\beq{e6}
\Delta \ub_\qb(t)  = \int_0^tdt'\; \vb_\qb(t')
\; .\eeq
Equations of motion (EOM) of displacement correlation functions thus follow from the familiar results on ${\bf K}(\qb,t)$ [14] based on the Zwanzig-Mori  projection operator formalism. We consider ${\bf C }(\qb,t)$, the tensor of collective mean-squared displacement differences defined by
\beq{e7}
{\bf C }(\qb,t) =   \langle \Delta \ub_\qb^*(t)\,\Delta \ub_\qb(t)  \rangle = 2 \int_0^t\!\!\! dt'(t\!-\!t'){\bf K}(\qb,t') .\eeq
Its EOM follow straightforwardly after overdamping as appropriate for colloidal dispersions:
\beq{e9}
{\bf C}(\qb,t)+ \frac{D_0 q^2}{k_BTn}\; \int_0^tdt'\; \tilde{\bf G}(\qb,t-t') \; {\bf C}(\qb,t') = 2D_0\, t\;  {\bf 1}\;\; ,
\eeq
with the short time diffusion coefficient $D_0$, and  stress kernels generalizing the (inverse) fluid isothermal compressibility $\kappa^T$ to finite wave vectors and frequencies:
$\tilde{\bf G}(\qb,t) = {\bf G}(\qb,t) + (1/\kappa_q^T) \hat{\qb}\hat{\qb}$. Here,  $\kappa^T_q=S_q/(k_BTn)$ is given by the equilibrium fluid structure factor.
The time-dependent stress kernels \beq{e8a}
{\bf G}(\qb,t)=(n/k_BT) \langle \sigb_{\qb}(t_{\pQ})^* \sigb_{\qb} \rangle\; ,
\eeq where $t_{\pQ}$ indicates  absence of conserved modes, are  built with the stress tensor elements containing the inter-particle forces (see Ch.~9.3, 9.4 in [14] and Ch.~3.3 in \cite{Goetze})
\beq{e8c}
\sigb_{\qb} = \frac{i}{q \sqrt{N}}\; \sum_{j=1}^N\;  \Fb_j
\;e^{i\rb_j \cdot {\bf q}}=\underbrace{\frac{-1}{\sqrt{N}}\; \sum_{j=1}^N\;
\frac{ \Fb_j(\rb_j \cdot \qb)}{q}}_{\sigb_0}+{\cal O}(q) \; .
\eeq
They reduce to the rheological stress auto-correlation functions in the limit of vanishing wave vector
\beq{e8b}
{\bf G}(t) = \lim_{q\to0} {\bf G}_\qb(t)  =  \frac{n}{k_BT}\; \langle  \sigb_0(t)^* \;\sigb_0  \rangle
\; .\eeq
These definitions and formally exact results will now be applied to glass, which is taken to be a non-ergodic state, where  the time dependent stress kernels take finite values at infinite time \cite{Goetze}:
\beq{e10}
{\bf G}(\qb,t\to\infty) \to {\bf G}_\infty(\qb)\;,
\eeq
which predicts from Eq.~(\ref{e9})
\beq{e11}
{\bf C}(\qb,t\to\infty) \to  {\bf C}_\infty(\qb) = 2 \frac{k_BT n}{q^2} \left({\bf G}_\infty(\qb)\right)^{-1}\; .
\eeq
Displacement differences stay below a finite limit for all times. This non-ergodic state is a solid one, and its displacement field can be obtained by integrating Eq.~\eqref{e6m} giving Eq.~(2) in the main text. The approximation $e^{ i \qb\cdot \rb_i(t)}=e^{ i \qb\cdot \bar{\rb}_i}+{\cal O}(\qb\cdot\ub_i(t))$ can be made.

Particles still locally move around their time-averaged positions, which can be measured by displacement functions defined as
\beq{e12}
\hat{\bf C}(\qb,t) = \langle  \ub_\qb(t)^*\;  \ub_\qb \rangle^{\rm glass}  .
\eeq The superscript 'glass' indicates that averaging is done in a restricted phase space set by the glassy state. As the so obtained displacement fluctuations are ergodic,  time and ensemble averages agree, and the $\hat{\bf C}(\qb,t)$ are auto-correlators [20]. Because of Eq.\eqref{e6m}, taking a time derivative of  $\hat{\bf C}(\qb,t)$ leads to velocity fluctuations which do not become non-ergodic at the glass transition owing to time-reversal symmetry \cite{Goetze}. This gives, again neglecting terms of order ${\cal O}(\qb\cdot\ub_i(t))$,
 \beq{e13}
\frac{\partial}{\partial t} \hat{\bf C}(\qb,t) =   - \frac 12\,\frac{\partial}{\partial t} {\bf C}(\qb,t)
\eeq
and thus the EOM of the displacement correlation functions $\hat{\bf C}(\qb,t)$  in glass can straightforwardly be obtained  from Eq.~\eqref{e9} except for an integration constant.    In order for $\hat{\bf C}(\qb,t)$ to approach zero at long times, this integration constant has to be chosen such that
\beq{e14}
\hat{\bf C}(\qb,t)  = \frac 12 \;\left(  {\bf C}_\infty(\qb) - {\bf C}(\qb,t) \right) .
\eeq
The equal time variance $\hat{\bf C}(\qb,t=0)= \frac 12   {\bf C}_\infty(\qb)$ follows, which is the equipartition theorem Eq.~(2) in the main text using Eq.~\eqref{e11} and definition \eqref{e12}.

\section{{\bf II:} Simulational details}
We simulated a binary mixture of hard discs undergoing Brownian motion using the  algorithm proposed by Scala et al.~\cite{Scala2007}.
The  system is made up of $N=1000$ particles, with a diameter ratio of small to big disks $d_s/d_b=0.7$ and equal number concentrations $x_s\!\!=\!x_b\!=\!1/2$ at a total packing fraction of $\varphi = \frac{\pi N}{4V} (x_s d_s^2 + d_b^2 x_b)$. A detailed analysis of the structural relaxation close to its glass transition can be found in Ref.~[23].

The dispersion relations and elastic moduli were obtained as explained in the letter from $10^4$ equally spaced snap-shots along one equilibrated simulation trajectory for times up to $\Delta t$.
Following the method of Alder et al.~\cite{Alder1970} we also determined the integrated time dependent shear modulus
\begin{align} \label{shearmod:shearviscosity}
& \eta_{xy}(t)
= \frac{1}{2k_B T V}  \frac{d}{dt} \left \langle \left( \sum \limits_{coll \in [0;t]} \Delta r^y_{ij} (t_c) \Delta p^x_{ij} (t_c)) \right)^2 \right \rangle .
\end{align}
Here the sum runs over all collisions up to time $t$ and $\Delta {\bf r}_{ij} (t_c)$ denotes the relative distance and $\Delta {\bf p}_{ij} (t_c)$ the momentum transfer of two particles at the collision at time $t_c$ \cite{Henrich2009}. The brackets $\langle...\rangle$ denote the  average over different simulation runs.
The integrated shear modulus was determined for $600$ independent, equilibrated initial configurations at $\varphi=0.81$ and for $150$ for all other packing fractions. Equilibration was assumed when the correlation functions became independent on the waiting time. The differentiation  in Eq.~(\ref{shearmod:shearviscosity}) was done numerically, as was the second one to obtain  $G(t)$ from $\eta_{xy}(t)=\int_0^t\!\!dt' G(t')$. Finally, a seven point running average was performed on the data for $G(t)$.
The  Kohlrausch fit to the data points at $\varphi =0.80$ was performed in the intervall $[1.0;\,10^4] /(D_0n^2)$.

For each packing fraction, the shear moduli in Fig.~2 contain the statistical average of $100$ independent simulation runs with different trajectory lengths. We calculated the displacement auto-correlation functions (Eq.~(3)) and performed a fit with the function $A \cdot q^2$ for $q \lesssim 0.7$, from which we infer the shear moduli and the errobars in Fig.~2.